\documentclass[a4paper,fleqn]{cas-dc}

\usepackage[numbers]{natbib}
\usepackage{wasysym}
\newcommand{\be}{\begin{equation}}
\newcommand{\ee}{\end{equation}}
\newcommand{\nn}{\nonumber}

\def\tsc#1{\csdef{#1}{\textsc{\lowercase{#1}}\xspace}}
\tsc{WGM}
\tsc{QE}
\tsc{EP}
\tsc{PMS}
\tsc{BEC}
\tsc{DE}

\begin{document}
\let\WriteBookmarks\relax
\def\floatpagepagefraction{1}
\def\textpagefraction{.001}
\shorttitle{Black hole pair production}
\shortauthors{Ashoorioon and Jahani Poshteh}

\title [mode = title]{Black hole pair production on cosmic strings in the presence of a background magnetic field}

\author[1]{Amjad Ashoorioon}
\ead{amjad@ipm.ir}

\author[1]{Mohammad Bagher Jahani Poshteh}
\ead{mbj.poshteh@gmail.com}

\address[1]{School of Physics, Institute for Research in Fundamental Sciences (IPM), P.O. Box 19395-5531, Tehran, Iran}

\begin{abstract}
We investigate the pair creation of magnetically charged black holes on a cosmic string in the presence of a background magnetic field. The string may either break or fray to produce a pair of accelerating black holes described by Ernst metric. By using the instanton action we obtain the rate of such production. For large values of background magnetic field the production of large black holes is probable. Comparing our results with the case of black hole pair creation in magnetic field with no string, we show that fraying/breaking of the cosmic string can substantially enhance the production rate. We also obtain the result of monopole-antimonopole pair production on cosmic string in an external magnetic field, using the WKB approximation and compare it with the black hole results. We also provide a heuristic study of black hole pair creation on cosmic string in the presence of a magnetic field in the de Sitter background, although the analog of Ernst metric with a cosmological constant is lacking. Like a cosmic string and/or a background magnetic field, a positive cosmological constant increases the production rate.
\end{abstract}


\maketitle

\section{Introduction}

Black hole pair creation has been an interesting subject, in the past few decades, partly because it leads to a deeper understanding of the black hole entropy~\cite{garfinkle1991,garfinkle1994,hawking1995,booth1998}. In fact, it has been shown in~\cite{garfinkle1994} that the pair production amplitude contains a factor of $e^{S_{bh}}$, where $S_{bh}$ denotes the black hole entropy. This is consistent with the view that $e^{S_{bh}}$ measures the number of black hole (internal or surface) states.

Pair creation through quantum mechanical tunneling effect can also explain the production of black holes less massive than a few Solar masses. Due to fermionic degeneracy pressure, gravitational collapse cannot be responsible for the production of these small black holes. A recent upper bound on the mass of cold spherical neutron stars is obtained to be $2.3\, M_{\astrosun}$, obtained using the results of GW170817, where it has been shown that neutron stars more massive than this limit would gravitationally collapse into a black hole ~\cite{shibata2019}.

A possible mechanism to produce black hole pairs is through breaking of a cosmic string~\cite{hr} (see also~\cite{achucarro1995}). The string could also ``fray'' to create black hole pair~\cite{eardley1995}. When there is no background magnetic field, it has been shown in~\cite{ashoorioon2014} that the tension of the strut between the two black holes, $\mu_4$, must be smaller than the tension of the string, $\mu_3$, from each black hole to infinity. Given the observational constraint on the tension, $\mu\leq 2\times 10^{-7}$~\cite{seljak2006}, either case of breaking or fraying of the cosmic string to produce black holes of mass larger than the Planck mass is rare~\cite{hr,eardley1995}.

On the other hand, pair creation of black holes in a background magnetic field, with no string, has been widely studied in the literature~\cite{garfinkle1991,dowker1994,hawking1995}. Using the instanton action, the rate of this process is found in the semiclassical approximation. As we show explicitly in this paper, for observationally viable values of the intergalactic magnetic field, the production rate of black holes of mass larger than the Planck mass is negligibly small.

In this paper we investigate the pair production of magnetically charged black holes on a cosmic string in the presence of a background magnetic field. To find the rate of this process, we use the action, $I$, of instanton interpolating between two following states. The first state is a spacetime with magnetic field and a cosmic string of tension $\mu_3$. We refer to this state as background spacetime. In the second state, which we refer to as physical spacetime, we have two black holes, each connected to a string of tension $\mu_3$ that runs to infinity. There may also be a strut between these two black holes with tension $\mu_4$. This configuration is in the presence of a magnetic field.

For a general value of the background magnetic field, we use numerical analysis and we find that the instanton action decreases as the background magnetic field increases. Since the pair production rate is given by $e^{-I}$, we find that for larger values of the background magnetic field the pair production is more probable.

We find an explicit relation for the action $I$ in the weak field limit. We compare this action, with that of the pair creation of black holes in a background magnetic field and no string. The setup we focus on is theoretical, however, we investigate the cosmological implications with real parameters. Using the cosmologically viable values for the tension of the cosmic string and the background magnetic field, we show that the presence of the string would significantly decrease the value of the action and, hence, increase the production rate.

For the sake of comparison with the black hole pair production, we study magnetically charged monopole-antimonopole pair production on a cosmic string in a background magnetic field. Here the configuration of physical spacetime is like that of the black hole case with black holes replaced by magnetic monopoles. To obtain the action of monopole pair production we use the WKB approximation.

We also heuristically study how the presence of a positive cosmological constant affects the black hole pair creation rate in the presence of a background magnetic field and cosmic string. We find that the pair creation rate is higher in a de Sitter (dS) background. During the inflation, where the cosmological constant has its highest value, black hole pair production is the most probable.

The outline of the paper is as follows. In the next section we review the Ernst metric which describes a pair of black holes connected to semi-infinite cosmic strings, in a background magnetic field. We study the production rate of such pairs in Sec.~\ref{sec:rate} by using the instanton action. The weak field limit of this action is studied in Sec.~\ref{sec:wf} where we provide some examples to show how the presence of a cosmic string would decrease the action. In Sec.~\ref{sec:mp} we study the pair production of monopole-antimonopole on a cosmic string in a background magnetic field.  A heuristic derivation of the action in a dS background is presented in Sec.~\ref{sec:ds}. We conclude our paper in Sec.~\ref{sec:con}. We use the units in which $G=c=\hbar=\epsilon_0=1$.

\section{Black hole pair on a cosmic string}\label{sec:ernst}

Before the pair creation, the spacetime consists of a cosmic string with tension $\mu_3$ in a magnetic field. We call this state, the background spacetime and it will be described by Melvin solution~\cite{melvin1964}
\begin{align}
	ds^2&=\bar{\Gamma}^2\left[-dt^2+dz^2+d\rho^2\right]+\bar{\Gamma}^{-2}\rho^2d\bar{\phi}^2, \label{melvin}\\
	\bar{\Gamma}&=1+\frac{1}{4}\hat{B}_{M}^2\rho^2, \nn
\end{align}
where $\hat{B}_{M}$ denotes the magnetic field, and the gauge potential is given by $2\bar{\Gamma}A_{\bar{\phi}}=\hat{B}_{M}\rho^2$. The cosmic string results in a deficit angle $\delta_3=8\pi\mu_3$ in the azimuthal coordinate, so $\Delta\bar{\phi}=2\pi-\delta_3$.

After the pair creation, the spacetime, which now we refer to as physical spacetime, will be given by the Ernst metric~\cite{ernst1976}
\begin{align}
	ds^2&=r^2\Gamma^2\left[G(y)dt^2-G^{-1}(y)dy^2+G^{-1}(x)dx^2\right] \nn\\ &+\frac{r^2G(x)}{\Gamma^2}d\phi^2, \label{ernst}\\
	A_{\phi}&=-\frac{2}{B\Gamma}\left(1+\frac{1}{2}Bqx\right)+k, \nn
\end{align}
where $r=A^{-1}(x-y)^{-1}$ represents the radial coordinate~\cite{kinnersley1970}, and
\begin{align}
	\Gamma&=\left(1+\frac{1}{2}B q x\right)^2+\frac{1}{4}r^2B^2G(x),\\
	G(\xi)&=(1+r_{-}A\xi)(1-\xi^2-r_{+}A\xi^3). \nn
\end{align}
The parameter $q^2=r_{-}r_{+}$ is related to the charge of the black hole and the mass of the black hole is given by $m=(r_{-}+r_{+})/2$. This solution describes two charged black holes accelerating from each other. The acceleration is given by the parameter $A$ with its inverse representing the typical distance between the two black holes at the time of their creation~\cite{dias2004}.

The physically interesting case is the one in which the function $G(\xi)$ has four real roots. One root is given by $\xi_1=-1/(r_{-}A)$. To have three other real roots we need to restrict the parameters so that $r_{+}A\leq\sqrt{4/27}$, which means that the black holes should be small and/or they should be far apart from each other. For $r_{+}A\ll 1$ one finds the remaining roots as follows
\begin{align}
	\xi_2&=-\frac{1}{r_{+}A}+r_{+}A+\cdots,\label{roots}\\
	\xi_3&=-1-\frac{r_{+}A}{2}+\cdots, \nn\\
	\xi_4&=1-\frac{r_{+}A}{2}+\cdots. \nn
\end{align}

To obtain the correct signature of the metric (\ref{ernst}), we require $\xi_3\leq x \leq \xi_4$ and $-\infty<y\leq x$. The surfaces $y = \xi_1$, $\xi_2$ , $\xi_3$ correspond to black hole inner horizon, event horizon, and acceleration horizon, respectively. Also $x=\xi_3$ and $x=\xi_4$ are axes that point to spatial infinity and to the other black hole, respectively. The parameter $k$, in the gauge potential of Eq. (\ref{ernst}), will be chosen so as to confine the Dirac string singularities to the $x=\xi_4$ axis. It should also be mentioned that the parameter $B$ in the Ernst metric is related to the physical magnetic field $\hat{B}$ by~\cite{dowker1994}
\be
\hat{B}=\frac{B\,G'(\xi_3)}{2L^{3/2}}, \label{physb}
\ee
where $L=\Gamma(\xi_3)$. It could be easily shown that in the weak field limit $\hat{B}=B$ to the leading order of background magnetic field. Also, the physical magnetic charge of the black hole is defined by $\hat{q} = 1/(4\pi)\int F$, where $F$ denotes the Maxwell field~\cite{dowker1994}. Integration over a two sphere surrounding the black hole yields
\be
\hat{q} = q \frac{\Delta\phi\left(\xi_4-\xi_3\right)}{4\pi L^{1/2}\left(1+\frac{1}{2}B q \xi_4\right)}, \label{eqn:charge}
\ee
where $\Delta\phi$ is the period of the azimuthal coordinate.

Consider the angular part of the Ernst metric (\ref{ernst})
\be
d\Phi^2=r^2\left[\frac{\Gamma^2}{G(x)}dx^2+\frac{G(x)}{\Gamma^2}d\phi^2\right].
\ee
Using the change of variable $\theta=\int_{\xi_i}^{x}\Gamma G^{-1/2}(x)dx$, near the poles $x=\xi_i$ ($i=3,4$), we obtain
\be
d\Phi^2=r^2\left[d\theta^2+\frac{G'^2(\xi_i)}{4\Gamma^4(\xi_i)}\theta^2d\phi^2\right].
\ee
We are interested in a configuration in which each black hole is connected to a string with deficit angle $\delta_3$ that runs to infinity and a strut with deficit angle $\delta_4$ that goes to the other black hole. In such case, one can show that the following relation holds near the poles
\be
\frac{|G'(\xi_i)|}{2\Gamma^2(\xi_i)}\Delta\phi=2\pi-\delta_i, \qquad i=3,4. \label{phiatpoles}
\ee
Now we write the period of the azimuthal coordinate in terms of $\xi_3$
\be
\Delta\phi=\frac{2L^2}{G'(\xi_3)}(2\pi-\delta_3)\,. \label{deltaphi}
\ee
It is easy to show that $G'(\xi_3)>0$. By using Eqs. (\ref{phiatpoles}) and (\ref{deltaphi}), we find
\be
\left|\frac{G'(\xi_4)}{G'(\xi_3)}\right|\frac{L^2}{\Gamma^2(\xi_4)}=\frac{2\pi-\delta_4}{2\pi-\delta_3}. \label{wbnewton}
\ee
This relation would lead to Newton's law in the weak field limit $Bq\ll1$. In observationally important case we have $\delta\ll1$. We are also working in the limit which the size of the black hole is small compared to the typical distance between them (the so called small black hole limit). Then Eqs. (\ref{roots}) and (\ref{wbnewton}) would result in
\be
mA=\left(\mu_3-\mu_4\right)+Bq. \label{newton}
\ee

Now consider the first two terms of the Ernst metric (\ref{ernst}). By Wick rotating the time coordinate $t=i\tau$ and following a similar calculations as above, we find that the period of the imaginary time at the surfaces $y=\xi_i$ ($i=2,3$) is $\Delta\tau=4\pi/|G'(\xi_i)|$. For the case of extreme black hole, $\xi_1=\xi_2$, the surface $y=\xi_2$ is infinitely far away from any point in the Euclidean sector of the metric~\cite{hawking1995}. As a result, one does not worry about the periodicity of $\tau$ at this surface (In this case we take $\Delta\tau=4\pi/G'(\xi_3)$). For the non-extreme case we request that $y=\xi_2$ and $y=\xi_3$ have the same periodicity of $\tau$, or equivalently, have the same temperature $1/\beta$. So
\be
\Delta\tau=\beta=\frac{4\pi}{G'(\xi_3)}=-\frac{4\pi}{G'(\xi_2)},
\ee
which leads to the following relation for the non-extremal case
\be
\xi_4-\xi_3=\xi_2-\xi_1. \label{nerelation}
\ee

\section{Production rate of black hole pair on a cosmic string in the presence of a background magnetic field}\label{sec:rate}

In this section we are going to obtain the rate of black hole pair production on cosmic string in the presence of a background magnetic field. To do so, we use the action of the instanton that interpolates between background and physical spacetimes. Based on the reasoning of~\cite{hr,hawking1996}, this action is equivalent to the Euclidean action
\footnote{We note that in Eqs. (\ref{neaction}) and (\ref{eaction}) all the surface terms have been taken into account~\cite{hawking1996}. The discussion of~\cite{hawking1996} applies to all spacetimes that can be foliated by complete, nonintersecting spacelike surfaces. In derivation of this formulae one needs to keep in mind that, since the Lorentzian action diverges for spatially non-compact geometries, one needs a reference background. One then fixes a boundary near infinity and requires that the metric and matter fields of the physical spacetime induce the same fields on the boundary as those of the background spacetime induce. Also, it is assumed that the periodicity of time coordinate $\beta$ be the same in physical and background spacetimes.}
\be
I=\beta H-\frac{1}{4}\Delta\mathcal{A}, \label{eaction}
\ee
in the extremal black hole case, and
\be
I=\beta H-\frac{1}{4}\left(\Delta\mathcal{A}+A_{bh}\right), \label{neaction}
\ee
in the non-extremal case.
\footnote{The extra term in Eq. (\ref{neaction}) relative to Eq. (\ref{eaction}) shows that the production rate in non-extreme case is enhanced by $e^{S_{bh}}$, where $S_{bh}=A_{bh}/4$ is the entropy of the non-extremal black hole, as the entropy of an extremal black hole is zero~\cite{hawking1995}. See also~\cite{emparan1995:2} to find out why only the entropy of one of the black holes enter the equations.}
Here $\Delta\mathcal{A}$ is the difference between the area of the acceleration horizon in physical and background spacetimes. $A_{bh}$ is the area of the black hole, and the Hamiltonian $H$ is given by~\cite{hawking1996}
\be
H=\int_{\Sigma}N\mathcal{H}-\frac{1}{8\pi}\int_{S_{\infty}^2}N\left(^2K-{}^2K_0\right), \label{hamilton}
\ee
where $N$ is the lapse function and $\Sigma$ is the constant time hypersurface. $\mathcal{H}$ is the Hamiltonian constraint which vanishes on the solution. $S_{\infty}^2$ is the boundary of $\Sigma$ and $^2K$ and $^2K_0$ are the extrinsic curvature of this boundary in physical and background spacetimes, respectively. Note that $S_{\infty}^2$ has been chosen so as to match in physical and background spacetimes. This way, a calculation similar to that of~\cite{hawking1995} would result in the conclusion that the second integral in Eq. (\ref{hamilton}) also vanishes.

Now, we are going to find the area of the black hole horizon, $A_{bh}$, which appears in the action (\ref{neaction}) for non-extremal black holes. The event horizon is at $y=\xi_2$ in the Ernst metric (\ref{ernst}). We find, by using (\ref{deltaphi}),
\begin{align}
	A_{bh}&=\int_{y=\xi_2}\sqrt{g_{xx}g_{\phi\phi}}dxd\phi \nn\\ &=\frac{2L^2}{A^2G'(\xi_3)}\left(2\pi-\delta_3\right)\frac{\xi_4-\xi_3}{\left(\xi_3-\xi_2\right)\left(\xi_4-\xi_2\right)}. \label{bharea}
\end{align}

Now, we proceed to find the area of the acceleration horizons. Recalling the definition of the radial coordinate, $r=A^{-1}(x-y)^{-1}$, and the restrictions on $x$ and $y$, we note that equalities $x = y = \xi_3$ show the spatial infinity. To find the area of the acceleration horizon in the Ernst metric, which is located at $y=\xi_3$, we integrate from $x=\xi_4$ to $x=\xi_3+\epsilon$, and will take the limit $\epsilon\rightarrow0$ at the end. We would find
\begin{align}
	\mathcal{A}&=\int_{y=\xi_3}\sqrt{g_{xx}g_{\phi\phi}}dxd\phi\nn\\ &=\frac{2L^2}{A^2G'(\xi_3)}\left(2\pi-\delta_3\right)\left(\frac{1}{\epsilon}-\frac{1}{\xi_4-\xi_3}\right). \label{ernstacch}
\end{align}
The area of the acceleration horizon in the Melvin metric (\ref{melvin}) is
\be
\bar{\mathcal{A}}=\int\rho d\rho d\bar{\phi}=\frac{2\pi-\delta_3}{2}\bar{\rho}^2. \label{melvinacch}
\ee
By requiring that the proper length of the boundary and the integral of the gauge potential around the boundary be the same in Ernst and Melvin metrics, it has been found in~\cite{hawking1995} that
\be
\bar{\rho}^2=\frac{4L^2}{A^2G'(\xi_3)\epsilon}\left(1+\frac{G''(\xi_3)}{4G'(\xi_3)}\epsilon\right)^2.
\ee
Substituting $\bar{\rho}$ from the above equation into Eq. (\ref{melvinacch}), we obtain
\be
\bar{\mathcal{A}}\simeq\frac{2L^2}{A^2G'(\xi_3)}\left(2\pi-\delta_3\right)\left(\frac{1}{\epsilon}+\frac{G''(\xi_3)}{2G'(\xi_3)}\right). \label{melvinacch2}
\ee
To find the second term in the above equation we note that one could write the function $G(\xi)$ as
\be
G(\xi)=-(r_-A)(r_+A)(\xi-\xi_1)(\xi-\xi_2)(\xi-\xi_3)(\xi-\xi_4).
\ee
By using Eqs. (\ref{ernstacch}) and (\ref{melvinacch2}) we find that
\begin{align}
	\Delta\mathcal{A}=\mathcal{A}-\bar{\mathcal{A}}&=-\frac{2L^2}{A^2G'(\xi_3)}\left(2\pi-\delta_3\right) \nn\\
	&\times\left(\frac{1}{\xi_3-\xi_1}-\frac{1}{\xi_3-\xi_2}\right). \label{acch}
\end{align}
By substituting this equation along with Eq. (\ref{bharea}) into Eq. (\ref{neaction}), and using the non-extremality relation (\ref{nerelation}), we obtain
\be
I=\frac{L^2}{A^2G'(\xi_3)}\left(\frac{2\pi-\delta_3}{\xi_3-\xi_1}\right). \label{action}
\ee
The same equation would be found for the action of extremal black hole, by using Eqs. (\ref{eaction}) and (\ref{acch}) and the relation $\xi_2=\xi_1$.

Here we would like to study the action (\ref{action}) for the case in which $q = m$ (or equivalently $r_+ = r_-$). We have restricted the black holes to be small on the scale set by the acceleration. That is, we are considering the case in which $r_+A\ll1$. By using this condition, we could show from (\ref{physb}) that
\be
\hat{B}\simeq\frac{8B}{(2-Bm)^{3}}. \label{physb:genb}
\ee
We see that $\hat{B}$ goes from zero to infinity as $B$ changes from zero to $2/m$.

\begin{figure}[htp]
	\centering
	\includegraphics[width=0.45\textwidth]{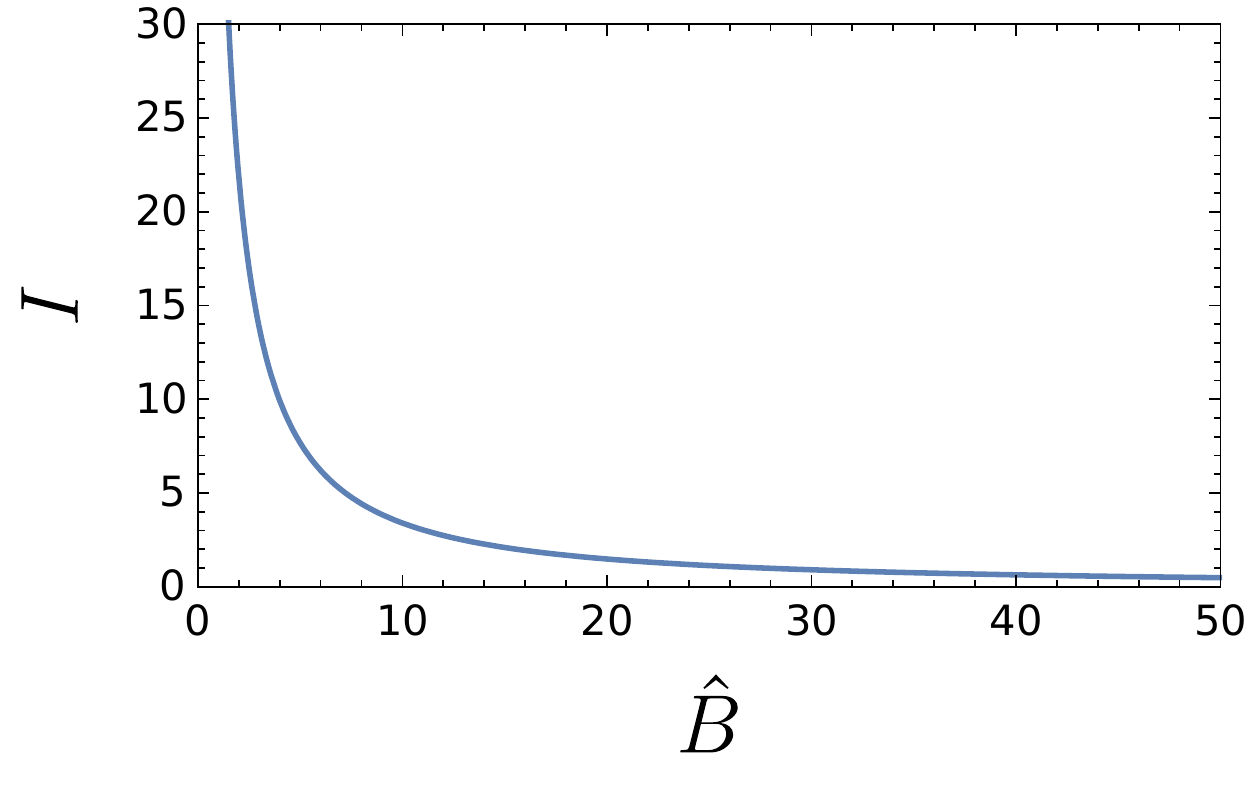}
	\caption{The action as a function of the physical magnetic field for the $q = m$ case. We see that the action decreases by increasing the physical magnetic field $\hat{B}$. We have set $m\,(=q)=10$, $\mu_3 = 2\times 10^{-7}$, and $\mu_4 = 10^{-7}$.}
	\label{fig:action}
\end{figure}

On the other hand, for the case $q = m$, Eq. (\ref{action}) could be written as
\be
I\simeq\frac{m(B m-2)^4}{32A}\left(2\pi-\delta_3\right), \label{action:genb}
\ee
where the parameter $A$ has to satisfy Eq. (\ref{wbnewton}). We have not been able to solve Eq. (\ref{wbnewton}) analytically, to find $A$ for a general value of $B$. However by using numerical methods, we have used Eqs. (\ref{physb:genb}), (\ref{action:genb}), and (\ref{wbnewton}) to plot the action as function of physical magnetic field $\hat{B}$ for the $q = m$ case in Fig.~\ref{fig:action}. It is obvious from this figure that the instanton action decreases as the physical magnetic field increases. Therefore, $e^{-I}\rightarrow1$ as $\hat{B}\rightarrow\infty$, which means that the black hole pair production rate increases by increasing the magnetic field.

\section{Weak field limit of the action}\label{sec:wf}

We see from Fig.~\ref{fig:action} that the action for the pair production of black holes of masses larger than the Planck mass is of order unity, only when the physical magnetic field is about 10 or larger. This is much larger than the value of intragalactic magnetic field, which, nowadays, is less than about $10^{-63}$ in natural units~\cite{ashoorioon2005}. It motivated us to study the weak field limit of the action (\ref{action}).

In the weak field limit, the physical magnetic field (\ref{physb}) and charge (\ref{eqn:charge}) can be approximated by
\begin{align}
	\hat{B}&\simeq B\left[1-\frac{1}{2}Bq-2\left(\mu_3-\mu_4\right)\right], \label{eqn:physb:emp}\\
	\hat{q}&\simeq q\left(1-4\mu_3\right)\left[1+2\left(\mu_3-\mu_4\right)\right],
\end{align}
where we have used Eq. (\ref{deltaphi}). These expressions reduce to the ones derived in ~\cite{emparan1995} when $\mu_3=0$. Using these equations, along with the Newton's law (\ref{newton}) and the non-extremality relation (\ref{nerelation}), one can write all the parameters in terms of $\hat{B}$, $\hat{q}$, $\mu_3$, and $\mu_4$
\begin{align}
	r_{\pm}&\simeq \hat{Q}\left[1-2\left(\mu_3-\mu_4\right)\pm\left(\hat{B}\hat{Q}+\mu_3-\mu_4\right)\right], \label{eqn:para:r}\\
	q&\simeq\hat{Q}\left[1-2\left(\mu_3-\mu_4\right)\right], \label{eqn:para:q}\\
	A&\simeq\hat{B}\left[1+\frac{1}{2}\hat{B}\hat{Q}+\frac{\mu_3-\mu_4}{\hat{B}\hat{Q}}\left(1+2\hat{B}\hat{Q}\right)\right], \label{eqn:para:A}
\end{align}
where $\hat{Q}=\hat{q}\left(1+4\mu_3\right)$. Again see similar expressions in ~\cite{emparan1995}, when $\mu_3=0$.

By substituting the parameters (\ref{eqn:para:r})-(\ref{eqn:para:A}) into Eq. (\ref{action}) we find the weak field approximation of the action as
\be
I\simeq\pi\hat{Q}^2\left(1-4\mu_3\right)\left[\frac{1}{\hat{B}\hat{Q}}-\frac{1}{2}-\frac{\mu_3-\mu_4}{\hat{B}^2\hat{Q}^2}\right]. \label{eqn:act:emp}
\ee
If we have no string, $\mu_3 = \mu_4 = 0$, one obtains the result of~\cite{dowker1994}. Hence, Eq. (\ref{eqn:act:emp}) is also in agreement with Schwinger result~\cite{schwinger1951}.
\footnote{Eq. \eqref{eqn:act:emp} agrees to the first three terms of the action obtained in~\cite{emparan1995}. However, if we continue our expansion \eqref{eqn:act:emp}, the fourth term would be $-\pi\hat{Q}\left(1-4\mu_3\right)\left(\mu_3-\mu_4\right)/\hat{B}$ which differs from the fourth term of~\cite{emparan1995} by a factor of 2. We only consider the first three terms of the action \eqref{eqn:act:emp} when we are comparing black hole pair production action to that of monopoles in the next section.}

Now we study the action in the case where $q=m$ and we are also in the weak field limit. In this case, the action (\ref{action}) can be written as
\be
I\simeq \frac{\pi m (1-2Bm)}{A}\left(1-4\mu_3\right). \label{action:smallb}
\ee
By using Eqs. (\ref{newton}) and (\ref{eqn:physb:emp}) one can rewrite the above equation as
\be
I\simeq \frac{\pi m^2 \left[1-2\left(J+\mu_3-\mu_4\right)\right]\left(1-4\mu_3\right)}{J-1+\left[1+2\left(\mu_3-\mu_4\right)\right]\left(\mu_3-\mu_4\right)}, \label{ours}
\ee
where the auxiliary function $J$ is
\be
J=\sqrt{\left[1-2\left(\mu_3-\mu_4\right)\right]\left[1-2\left(\hat{B}m+\mu_3-\mu_4\right)\right]}. \nn
\ee

Let us see how the presence of the cosmic string would change the rate of the production. Suppose that a pair of black holes, each of mass $m=10$, pop up in the background magnetic field $\hat{B}=10^{-63}$ which is the today's value of the intragalactic magnetic field. Substituting these values into Eq. (\ref{ours}) with $\mu_3 = \mu_4 = 0$, we find $I\simeq 3.141592\times 10^{64}$ for the action of this pair creation. Let us compare this with the case where $\hat{B}=10^{-63}$, $m=10$, and $\mu_3 = \mu_4 =2\times 10^{-7}$ i.e.~when the pair creation of black holes occurs on a cosmic string with uniform tension. In this case we find $I\simeq 3.141590 \times 10^{64}$, which is slightly smaller than when black holes pair is created from vacuum. This is due to the factor $1-4\mu_3$ in Eq. \eqref{ours}. This implies that the presence of the string, even if it does not fray or break, reduces the action, although for the cosmic strings with observationally relevant values of tension, the change in the action will be quite small.

Now let us assume that we have a string of tension $\mu_3=2\times 10^{-7}$ in the Melvin universe (recall that this is the upper limit on the tension of the cosmic string). Assume that the string frays and produces a pair of black holes with the same mass $m=10$. Let us take $\mu_4=10^{-7}$, for the tension of the strut between the two black holes. Also take the same value of the background magnetic field as above, $\hat{B}=10^{-63}$. Substituting these values into the action (\ref{ours}), we find $I\simeq 3.141590 \times 10^{9}$.
\footnote{We note that if we omit the factor $\left(1-4\mu_3\right)$ in Eq. (\ref{ours}), we would find $I\simeq 3.141592 \times 10^{9}$. Therefore, the contribution of this factor is very small compared to the changes of the action upon breaking/fraying of the cosmic string.}
This is smaller than the value of the action in the absence of the string, by 55 order of magnitude!

In the early universe, the background magnetic field can be much stronger. It is supposed to be of the order $\hat{B}=10^{-13}$ in natural units at the end of inflationary era \footnote{We have assumed that the observed intragalactic field, has been produced during inflation but its strength has reduced super-adiabatically, $B\propto a^{-2}$, since then.}. Using this value of the background magnetic field along with $\mu_3 = \mu_4 = 0$ and $m=10$, we find $I\simeq 3.141592 \times 10^{14}$,  which is smaller than the corresponding value in today magnetic ﬁeld strength by 50 orders of magnitude.

For the case of pair creation on the cosmic string in the early universe, we use $\mu_3=2\times 10^{-7}$, $\mu_4=10^{-7}$, $m=10$, and $\hat{B}=10^{-13}$. By substituting these values into Eq. (\ref{ours}), we find $I\simeq 3.141559 \times 10^{9}$. One notices that the value of the action for pair creation on a cosmic string at the end of the inflationary phase does not differ significantly from its current value, although without cosmic string, the action reduces significantly.

We find from Eq. (\ref{ours}) that for fixed values of the parameters, the action is minimized if the string breaks, i.e.~$\mu_4=0$. For $\hat{B}=10^{-63}$, $m=10$, $\mu_3=2\times 10^{-7}$, and $\mu_4=0$ we find $I\simeq 1.570795 \times 10^{9}$. On the other hand if the cosmic string does not fray/break, one finds the maximum value of the action for a given set of the parameters, even though we should stress that the obtained value in this case is slightly smaller than the value of the action when the pair creation of black holes takes place only in presence of background magnetic field without a cosmic string.

\begin{figure}[htp]
	\centering
	\includegraphics[width=0.45\textwidth]{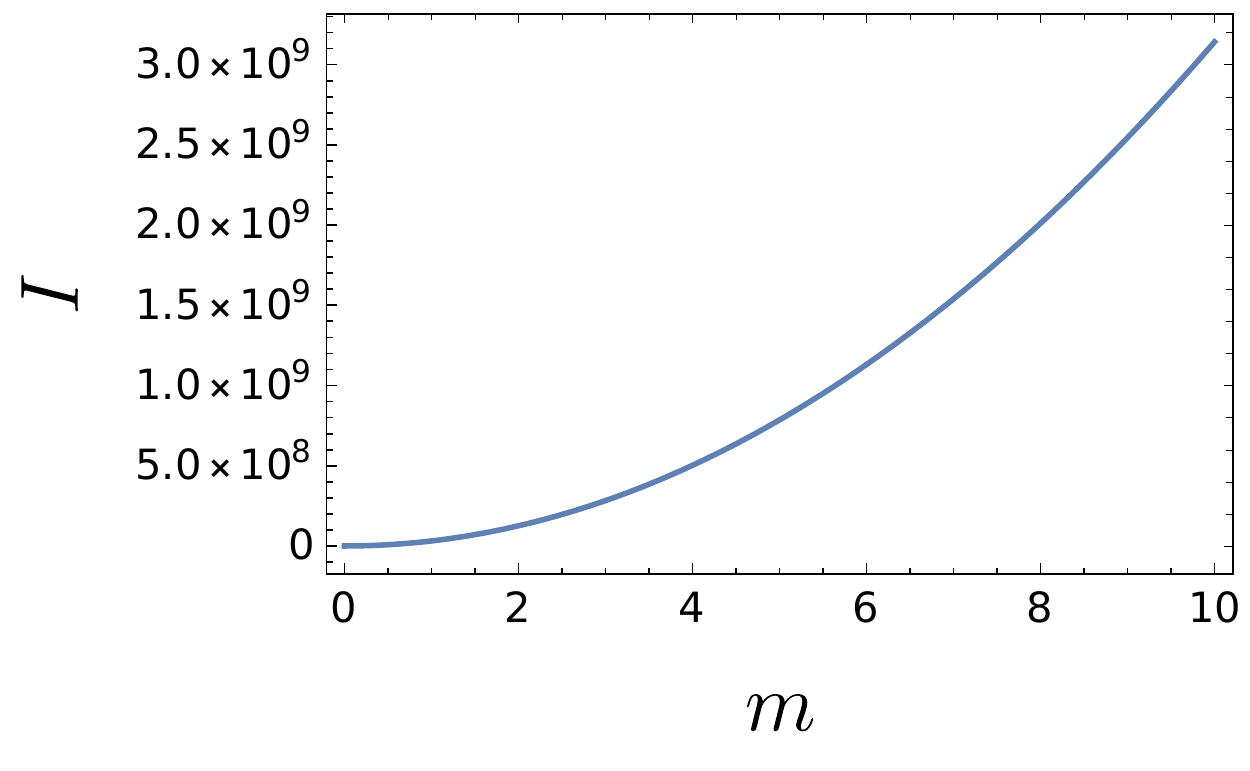}
	\caption{The action for the $q=m$ case in the weak field limit as a function of black hole mass. We see that the action increases by increasing the mass of the black holes. We have taken $\mu_3=2\times 10^{-7}$, $\mu_4=10^{-7}$, and $\hat{B}=10^{-63}$.}
	\label{fig:actionwf}
\end{figure}

We also see from Fig.~\ref{fig:actionwf} that for fixed $\mu_3$, $\mu_4$, and $\hat{B}$, the action (\ref{ours}) would increase by increasing the mass of the black holes. One concludes that the probability of pair producing more massive black holes is smaller.

\section{Monopole pair production on cosmic strings in a background magnetic field}\label{sec:mp}

Monopoles are hypothetical particles with quantized magnetic charge~\cite{dirac1931}. It has been shown that they can be produced in theories with certain non-Abelian gauge fields~\cite{t1974}. In this section we study the production rate of magnetically charged monopole-antimonopole pair on cosmic string in the presence of a background magnetic field.

In the absence of a background magnetic field, the semiclassical probability for a cosmic string to break producing a monopole-antimonopole pair has been obtained and their cosmological evolution has been studied~\cite{vilenkin1982}. The production rate of monopole pairs in a background magnetic field has also been studied in~\cite{affleck1982}. Recently, the rate of monopole pair production has been investigated in thermal baths~\cite{gould2017} and strong background magnetic fields~\cite{rajantie2019,ho2020}. Investigation of monopole pair production on cosmic string in a background magnetic field is missing in the literature.

Here we use WKB approximation, similar to that of~\cite{vilenkin1982}, to find the action of monopole pair production on cosmic string in an external magnetic field. We compare this result with the action of black hole pair production on a cosmic string in a background magnetic field.

Assume a cosmic string with tension $\mu_3$ on an axis $z$. Assume also a background magnetic field $\hat{B}$ parallel to this axis. Suppose the string breaks and produces a monopole and an antimonopole each connected to a semi-infinite string. Because of the symmetry, we only consider the monopole in the following analyze and include the antimonopole rule by multiplying the monopole action by 2, in the end.

The monopole has mass $m$ and magnetic charge $\hat{q}$. The energy of the monopole as a function of $z$ is $m+\left(\mu_3+\hat{B}\hat{q}\right)z$. We note that if the string frays and we have strut of tension $\mu_4$ between the pair, then the force on monopole due to the string-strut system is $\mu_3-\mu_4$ and the energy would be
\be
E=m+\left(\mu_3-\mu_4+\hat{B}\hat{q}\right)z.
\ee
The additive constant has been chosen to make the monopole bounce at $z=0$, see \cite{vilenkin1982}. The action for the pair production is then
\be
I_{mon}\simeq 4 \int_{\frac{-m}{\mu_3-\mu_4+\hat{B}\hat{q}}}^{0}\sqrt{m^2-E^2}dz.
\ee
The lower bound of the integral has been chosen so as to make $E\geq 0$ throughout the integration domain. Integrating we find
\be
I_{mon}\simeq \pi m^2 \left(\frac{1}{\hat{B}\hat{q}}-\frac{\mu_3-\mu_4}{\hat{B}^2\hat{q}^2}\right). \label{eqn:mnac1}
\ee

It has been suggested in~\cite{dowker1994} that, regarding the production rate, monopoles behave more like extremal black holes than non-extremal ones. This suggestion is based on the calculations of~\cite{garfinkle1994} which shows that the action of the instanton describing monopole pair production is greater than the action of non-extreme black holes by the black hole entropy. On the other hand, it was found in~\cite{dowker1994} that the instanton action for the pair production of extremal black holes is larger than the action for the case of non-extreme black holes again by the black hole entropy. For the extreme black holes we have $m=\hat{q}$. Thus if we regard monopoles as extreme black holes we can rewrite Eq. \eqref{eqn:mnac1} as
\be
I_{mon}\simeq \pi \hat{q}^2 \left(\frac{1}{\hat{B}\hat{q}}-\frac{\mu_3-\mu_4}{\hat{B}^2\hat{q}^2}\right). \label{eqn:mnac2}
\ee

Eq. \eqref{eqn:act:emp} is the action for pair production of (slightly) non-extremal black holes, and, in $\mu_3\ll 1$ limit, can be written as the following
\be
I\simeq \pi \hat{q}^2 \left(\frac{1}{\hat{B}\hat{q}}-\frac{\mu_3-\mu_4}{\hat{B}^2\hat{q}^2}-\frac{1}{2}\right).
\ee
Comparing this equation with Eq. \eqref{eqn:mnac2}, we find that the action obtained by WKB approximation is greater than the action of non-extremal black holes by half the Bekenstein-Hawking entropy $\pi\hat{q}^2/2=S_{bh}/2$. Similar discrepancies between the actions for non-extremal black hole pair creation, \cite{dowker1994}, the exact action for monopole pair creation, \cite{affleck1982}, and the WKB result for the monopole pair generation, all in a uniform magnetic background, is noticed. The difference is probably mainly due to the inadequacy of the WKB approach in obtaining the exact result. More on this and related issues remain to be studied.

\section{Heuristic derivation of the production rate in a de Sitter background}\label{sec:ds}

In this section we study black hole pair production on a cosmic string in the presence of a background magnetic field along with a positive cosmological constant $\Lambda$. The rate of black hole pair creation in dS background has been first studied in~\cite{mellor1989}, and further investigated in~\cite{mann1995}. It was shown that the action of the instanton that mediate between dS spacetime and dS spacetime with a pair of black holes, each of mass $m$, is $I = \pi m \sqrt{3/\Lambda}$. Later, in~\cite{dias2004:2}, a cosmic string was invoked in a dS background and the rate of black hole pair production has been studied.

To study the pair production of dS black holes in the presence of a background magnetic field, one needs a dS version of the Ernst metric. However such a solution does not exist~\cite{dias2003}. In fact, Ernst has used an Ehlers-Harrison type transformation~\cite{ehlers1959,harrison1968} to add the magnetic field to the C metric. But, applying this transformation to the cosmological C metric would not yield a new solution to the Einstein-Maxwell theory~\cite{dias2003}.

Nonetheless we can deal with this problem heuristically. One can think of a positive cosmological constant doing the same role as the cosmic string; in the sense that it provides the energy to materialize the pair of black holes and, also, provides the force to accelerate the black holes apart.

To find the acceleration caused by a positive cosmological constant, we recall that the Newtonian potential of dS spacetime is $\Lambda r^2/6$. The force per unit mass or the acceleration would then be $\Lambda r/3$, where $r$ is the dS radius given by $r=\sqrt{3/\Lambda}$. Therefore, in the weak field limit, we can modify the Newton's law (\ref{newton}) to be
\be
mA=\tilde{\mu}+Bq,
\ee
where $\tilde{\mu}=\mu_3-\mu_4+m\sqrt{\Lambda/3}$. Likewise, we can modify the actions (\ref{eqn:act:emp}) and (\ref{ours}) by replacing $\mu_3-\mu_4$ by $\tilde{\mu}$.

Consider, for instance the $q=m$ case in the weak field limit. The dS counterpart of the action (\ref{ours}) is
\be
I\simeq \frac{\pi m^2 \left[1-2\left(\tilde{J}+\tilde{\mu}\right)\right]\left(1-4\mu_3\right)}{\tilde{J}-1+\left(1+2\tilde{\mu}\right)\tilde{\mu}}, \label{ours:ds}
\ee
where
\be
\tilde{J}=\sqrt{\left(1-2\tilde{\mu}\right) \left[1-2\left(\hat{B}m+\tilde{\mu}\right)\right]}. \nn
\ee

In the limit $\Lambda\ll m^{-2}$, and $\hat{B},\, \mu_3,\, \mu_4 \rightarrow 0$, Eq. (\ref{ours:ds}) reduces to $I\simeq \pi m \sqrt{3/\Lambda}$, to leading order of $\Lambda$. This is consistent with the result of~\cite{mann1995} for the case of lukewarm black holes in which one has non-extreme black holes satisfying $q^2=m^2$~\cite{romans1992}.

We note here that if we modify the parameters (\ref{eqn:para:r})-(\ref{eqn:para:A}) to include the cosmological term and substitute them into Eq. (\ref{action}), then, the limit $\Lambda\ll m^{-2}$, and $\hat{B},\, \mu_3,\, \mu_4 \rightarrow 0$ would lead to $I\simeq \pi m \sqrt{3/\Lambda}$ as well. This is because in this limit we have $r_+\rightarrow r_-$, or equivalently $q\rightarrow m$. Therefore, again we are left with lukewarm black holes. This result shows that the lukewarm instanton is the only one available for non-extreme black holes, for which the conical singularities are eliminated at both $y=\xi_2$ and $y=\xi_3$ horizons in a dS background (see~\cite{mann1995} and~\cite{dias2004:2}).

Now let us study the action (\ref{ours:ds}) through some examples, considering the current bounds on the physical quantities. For $\hat{B}=10^{-63}$, $\mu_3=2\times 10^{-7}$, $\mu_4=10^{-7}$, $m=10$, and $\Lambda=10^{-122}$ we find $I\simeq 3.141590 \times 10^{9}$. On the other hand, if we have no string (i.e.~$\mu_3=\mu_4=0$) and keep the other parameters the same, we find $I\simeq 5.348755 \times 10^{62}$.

For the inflationary era, by taking $\hat{B}=0$, $\Lambda=10^{-12}$, $m=10$, and $\mu_3=2\times 10^{-7}$, and $\mu_4=10^{-7}$, we find $I\simeq 5.348751 \times 10^{7}$. For the no string case (i.e.~$\mu_3=\mu_4=0$) it is $I\simeq 5.441398 \times 10^{7}$, which is a bit larger. Although the value of the action for pair creation of black holes would not change significantly during inflation regardless of the presence of the cosmic string, the action is still considerably smaller in this era than the time it frays in presence of today's background magnetic field.

\section{Concluding remarks}\label{sec:con}

The rate of the black hole pair production (per unit time per unit length of the cosmic string) is given by $e^{-I}$. We see from Fig.~\ref{fig:action} that as the background magnetic field increases the black hole pair creation would be more probable. These black holes are accelerated away from each other after their production due to the force of the string and the background magnetic field.

We worked on a theoretical basis, however, we have presented our numerical examples with observationally viable values. Since the background magnetic field in the Universe is very small, we could safely use the weak field limit of the instanton action in cosmological applications. We have compared the pair creation of black holes in a background magnetic field (and no string) with that on a cosmic string in the presence of the background magnetic field. We have shown that the fraying/breaking of the cosmic string, even with the observational bound on its tension, would decrease the action significantly.

In~\cite{emparan1995}, the production rate of black hole pair, connected by a strut but with no string, was obtained in the Melvin universe. It was shown that this rate is suppressed relative to the case where no strut is present. Here we showed that if a cosmic string is present in the Melvin universe from the beginning, the pair production rate is enhanced.

In the early universe at the end of the inflationary era, the background magnetic field is expected to be much larger if they are produced during inflation. We find that the action of the pair production in no string case could be much smaller in that period. However if the pair production took place on a cosmic string, the value of the action at the end of the inflationary era and present day universe do not differ significantly.

We have studied the production rate of monopole-antimonopole pair as well. Here we have used WKB approximation to find the exponent of the rate. We have found that the rate of monopole pair production is suppressed relative to non-extremal black hole pair production by the factor $e^{S_{bh}/2}$. This is not consistent with the suggestion that monopoles behave like extremal black holes in the sense that their production is suppressed relative to non-extremal black hole production by the factor $e^{S_{bh}}$. There might be a problem with the WKB approach. There are various loop corrections with the same order as black hole entropy that have not been taken into account in WKB approximation~\cite{affleck1982,garfinkle1994}. This issue needs further investigations in future studies.

We have also presented a heuristic derivation of the production rate on cosmic string in the presence of a background magnetic field in dS spacetime. Like the cosmic string the presence of a positive cosmological constant decreases the action, hence increases the production rate. By using the action obtained in this way, we find that the pair creation is the most probable in the inflationary era. Contrary to the epoch after inflation, the black hole pair creation rate during inflation is nearly insensitive to whether it takes place on a cosmic string.

In our numerical example we have taken the mass of the black holes to be $m=10$ in Planck units. Such small black holes may tolerate the evaporation~\cite{macgibbon1987}. Besides, for the evaporation of the magnetically charged black holes, one needs copious magnetically charged monopoles to take away the charge of the black holes. This is not observationally acceptable as we do not see many (if any) monopoles around us. In GUT theories with magnetically charged monopoles with with mass $10^{16}~{\rm GeV}\lesssim \mu \lesssim  10^{19}~{\rm GeV}$, it has been shown that the black holes with mass $m\geq 150 \left(\frac{\mu}{10^{17}~{\rm GeV}}\right)^2$ can become stable roughly over the lifetime of the universe \cite{Hiscock:1983ag}. The temperature of the accelerating black holes is also somewhat subtle to define (see~\cite{appels2016,appels2017,appels2020}). More study on the thermodynamics of such black holes is needed.

\bibliographystyle{model1-num-names}
\bibliography{cas-refs}

\end{document}